\documentclass[runningheads]{llncs}
\usepackage[T1]{fontenc}
\usepackage{hyperref} 
\usepackage{graphicx}
\begin{document}
\title{Understanding Stain Separation Improves Cross-Scanner Adenocarcinoma Segmentation with Joint Multi-Task Learning}
\titlerunning{COSAS Segmentation Task 2 - Stain Separation with Multi-Task Learning}
%
\author{
    Ho Heon Kim\inst{1}\orcidID{0000-0001-7260-7504} \\
    Won Chan Jeong\inst{1}\orcidID{0009-0008-1931-5957} \\
    Young Shin Ko\inst{1}\orcidID{0000-0003-1319-4847} \\
    Young Jin Park\inst{1}\orcidID{0000-0002-0754-8103}
}
\authorrunning{
    HH Kim, and WC Jeong et al.
}
%

\institute{
    AI Research Center, Seegene Medical Foundation, Seoul, South Korea \\
    \email{
        \{hoheon0509, jeongwonchan53, noteasy, youngjpark\}@mf.seegene.com
    }
}
\maketitle              
\begin{abstract}
Digital pathology has made significant advances in tumor diagnosis and segmentation, but image variability due to differences in organs, tissue preparation, and acquisition - known as domain shift - limits the effectiveness of current algorithms. The COSAS (Cross-Organ and Cross-Scanner Adenocarcinoma Segmentation) challenge addresses this issue by improving the resilience of segmentation algorithms to domain shift, with Task 2 focusing on adenocarcinoma segmentation using a diverse dataset from six scanners, pushing the boundaries of  clinical diagnostics. Our approach employs unsupervised learning through stain separation within a multi-task learning framework using a multi-decoder autoencoder. This model isolates stain matrix and stain density, allowing it to handle color variation and improve generalization across scanners. We further enhanced the robustness of the model with a mixture of stain augmentation techniques and used a U-net architecture for segmentation. The novelty of our method lies in the use of stain separation within a multi-task learning framework, which effectively disentangles histological structures from color variations. This approach shows promise for improving segmentation accuracy and generalization across different histopathological stains, paving the way for more reliable diagnostic tools in digital pathology.

\keywords{Image Segmentation  \and Stain separation \and Multi-task learning.}
\end{abstract}
%
%
%


\section{Introduction}
\paragraph{}
The field of digital pathology has advanced significantly, especially in tumor diagnosis and segmentation. \cite{pathology} However, the effectiveness of current algorithms is often limited by the variability in digital pathology images, caused by differences in organs, tissue preparation, and image acquisition—a challenge known as domain-shift. \cite{variation} Addressing domain-shift is crucial for ensuring consistent performance of segmentation algorithms across different domains.

The COSAS  (Cross-Organ and Cross-Scanner Adenocarcinoma Segmentation) challenge  aims to develop and test strategies that enhance the resilience of computer-aided semantic segmentation to domain-shift, promoting reliable performance across various organs and scanners \cite{cosas}. This challenge is a significant step in advancing AI and ML for routine diagnostic use in clinical labs and is the first to provide a platform for evaluating domain adaptation methods on a multi-organ, multi-scanner dataset.
\
Specifically, Task 2 of the COSAS challenge focuses on assessing the generalization capabilities of algorithms in adenocarcinoma segmentation. It uses a diverse dataset of whole slide image patches of invasive breast carcinoma tissue from six different scanners. This task aims to push the boundaries of model development and promote robust solutions for real-world diagnostic applications in digital pathology.

\section{Methodology}
\subsection{Multi-task learning based on stain separation}
\paragraph{}
Motivated by unsupervised learning based on stain separation, our approach for joint multi-task learning leverages stain separation to isolate the stain matrix $W$ (stain color appearance) and stain map density $H$ \cite{moyes2020unsupervised}. Given Beer-Lambert transformed image $I \in R^{ m \times n}$, where $m$ is the number of channels, and $n$ is the number of pixels, the image, $I$, can be decomposed into $HW$. In this decomposition, the stain matrix $W \in R^{m \times r}$ represents the basis colors for each stain \cite{vahadane} with $r$ being the number of stains. The stain density $H \in R^{r \times n}$ represents the concentration of stains at each pixel. 

\begin{figure}
\includegraphics[width=\textwidth]{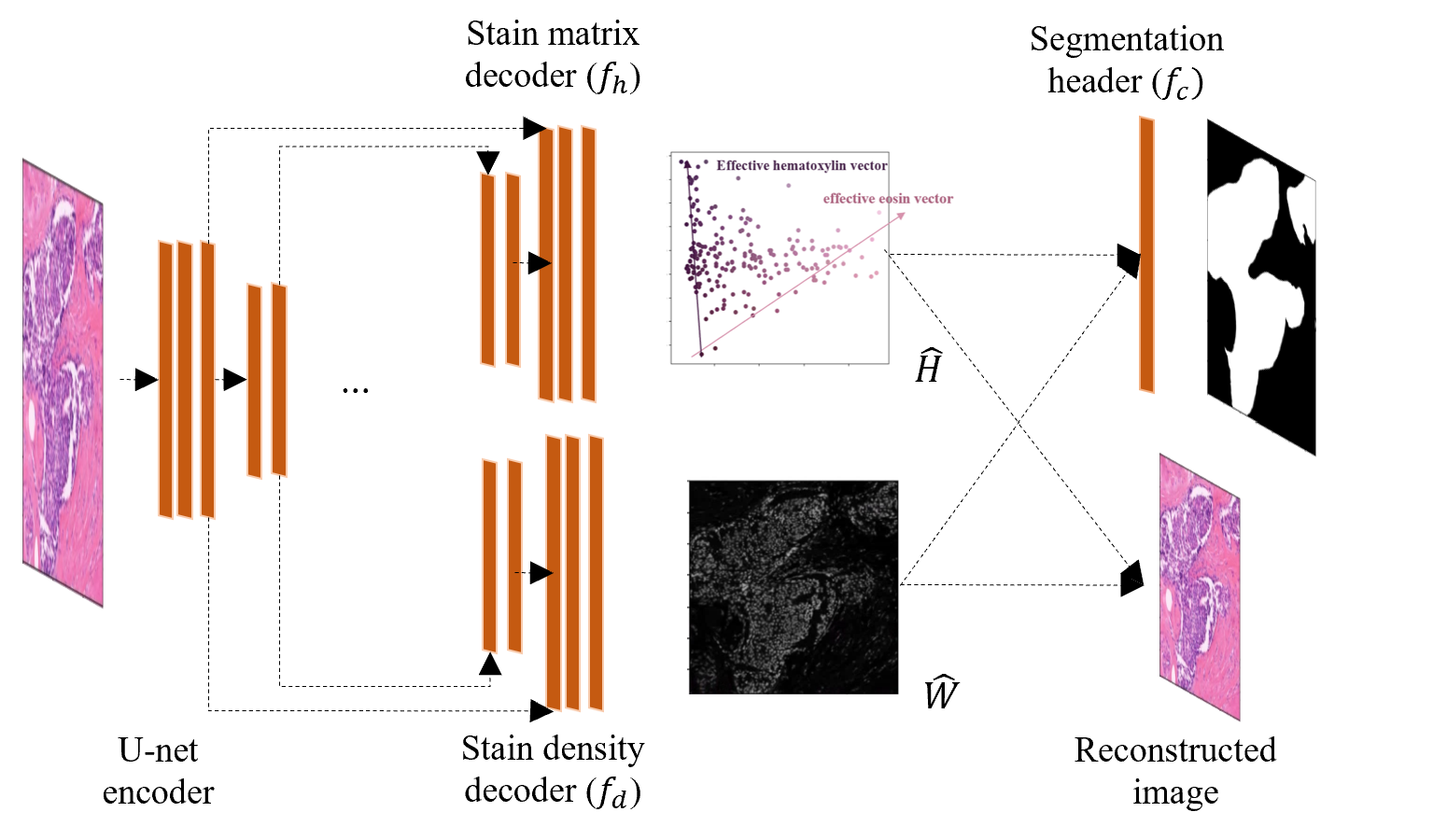}
\caption{Multi-decoder Unet architecture for joint multi-task learning} \label{fig1}
\end{figure}

By learning to separate these stain components, our model is designed to effectively capture histological structures from stain density and manage variations from the stain matrix, despite differences in scanner color bases. We implemented a multi-decoder AutoEncoder within a multi-task learning framework, where each decoder serves a specific function: 1) the stain matrix decoder ($f_{m}$) predicts the stain matrix ($\hat{W}$) to understand various colour styles at a pixel-wise level ($\hat{W} \in R^{n \times m \times r }$), addressing inter-stain variance among pixels \cite{vahadane}. 2) the stain density decoder estimates the predicted stain density ($\hat{H}$). A classification header  ($f_{c}$) is then attached to the model, leveraging feature maps from both decoders for segmentation (Fig. \ref{fig1}).

We hypothesized that if the stain matrix header ($f_{h}$) and stain density header ($f_{d}$) effectively learn histological information, their corresponding convolutional neural network (CNN) kernels would be well-trained. Consequently, these well-trained kernels would generate information feature maps, especially for H\&E stained images. To utilize these outputs, we concatenated the predicted stain density and stain matrix ($\hat{H}, \hat{W}$) and fed them into the classification header, yielding pixel-wise logits ($\hat{y}$) for segmentation:
\begin{equation}
    \hat{y} = f_{c}(\hat{H} \oplus \hat{W})
\end{equation}

For joint multi-task learning, we formulated the objective function ($\mathcal{L}_{total}$) as a weighted average of two loss functions: reconstruction error ($\mathcal{L}_{recon}$), calculated from elementwise multiplication of $\hat{H}$ and $\hat{W}$, and the pixel-wise classification error ($\mathcal{L}_{seg}$) for segmentation:
\begin{equation}
\mathcal{L}_{total} = \alpha \mathcal{L}_{recon} + \mathcal{L}_{seg}
\end{equation}
Here, $\alpha$ denotes the coefficient for the reconstruction error, balancing the contributions of the two loss functions. Code is available at: \url{https://github.com/4pygmalion/cosas}

\subsection{Mixture of stain augmentation}
To train a domain-generalized deep learning model, we created the realistic histopathologic image by mixing two stain augmentation methods: 1) RandStainNA \cite{randstainna} and 2) stain separation-based stain augmentation \cite{hoheon}. We mixed the augmentation probabilities by giving 0.25 to RandStainNA, and Stain separation based stain augmentation respectively. The stain separation-based augmentation technique uses SPCN (Structure-Preserving Color Normalization) to extract stain vectors and stain densities, modifying the stain vector's color basis through random distribution. This approach produces realistic H\&E stained images while preserving essential histological information, as SPCN only alters the stain matrix in terms of $H$.

\subsection{Training and inference configuration}
\paragraph{}
For the segmentation of adenocarcinoma, we employed a U-Net architecture built on a pretrained EfficientNet-B7 backbone derived from ImageNet1K \cite{unet,efficientnet}. The model training involved four-fold stratified cross-validation using data from three different scanners. In addition to stain augmentation, we applied random vertical and horizontal flips to enhance generalization. Furthermore, test-time augmentation was conducted by randomly rotating input images in 90° increments, creating four possible orientations (90°, 180°, 270°, and 360°). The COSAS score was calculated using the average of the Dice score and IoU (intersection over union).

\section{Results}
\subsection{Model performance}
Our model achieved an average COSAS metric of 0.846 in a 4-fold stratified cross-validation, with a Dice score of 0.887 and an IoU score of 0.805. In preliminary testing, the model achieved a COSAS score of 0.8775, but ultimately stabilized at 0.792 in the final evaluation. Sensitivity analysis on the weight coefficients of the loss function showed that a value of 0.3 for $\alpha$ provided the optimal balance as determined by grid search.

\section{Discussion}
The novelty of our methodology lies in the use of unsupervised learning for stain separation within a multi-task learning framework, enabling the model to effectively disentangle histological structures from color variations. By implementing a multi-head autoencoder that individually predicts stain matrix and stain density, our approach facilitates a more accurate understanding of color styles and tissue structures in heterogeneous scanner images. This integration into the classification header improves segmentation accuracy by exploiting rich histological information. Future work will aim to further optimize the stain augmentation techniques to improve generalization across even more diverse datasets, and to extend our approach to additional histopathological stain types beyond H\&E stains.

\section{Conclusion}
Stain separation, by isolating stain matrix and stain density through a multi-head autoencoder, enhances cross-scanner adenocarcinoma segmentation. This approach not only increases segmentation accuracy but also provides a robust solution to address domain shift in digital pathology.

\subsubsection{\discintname}
\begin{credits}
None.
\end{credits}


\begin{thebibliography}{8}

\bibitem{pathology}
Niazi, M. K. K., Parwani, A. V., \& Gurcan, M. N. (2019). Digital pathology and artificial intelligence. The Lancet Oncology, 20(5), e253–e261. https://doi.org/10.1016/S1470-2045(19)30154-8

\bibitem{variation}
Tiard, A., Wong, A., Ho, D. J., Wu, Y., Nof, E., Goh, A. C., ... \& Nadeem, S. (2022). Stain-invariant self supervised learning for histopathology image analysis. arXiv preprint arXiv:2211.07590.

\bibitem{cosas}
Grand Challenge: Cross-Organ and Cross-Scanner Adenocarcinoma Segmentation. (n.d.). Grand Challenge. https://cosas.grand-challenge.org/

\bibitem{moyes2020unsupervised}
Moyes, A., Zhang, K., Ji, M., Zhou, H., \& Crookes, D. (2020). Unsupervised deep learning for stain separation and artifact detection in histopathology images. In Medical Image Understanding and Analysis: 24th Annual Conference, MIUA 2020, Oxford, UK, July 15-17, 2020, Proceedings 24 (pp. 221-234). Springer International Publishing

\bibitem{vahadane}
Vahadane, A., Peng, T., Sethi, A., Albarqouni, S., Wang, L., Baust, M., ... \& Navab, N. (2016). Structure-preserving color normalization and sparse stain separation for histological images. IEEE transactions on medical imaging, 35(8), 1962-1971.

\bibitem{randstainna}
Shen, Y., Luo, Y., Shen, D., \& Ke, J. (2022, September). Randstainna: Learning stain-agnostic features from histology slides by bridging stain augmentation and normalization. In International Conference on Medical Image Computing and Computer-Assisted Intervention (pp. 212-221). Cham: Springer Nature Switzerland.

\bibitem{hoheon}
Kim, Ho Heon, et al. "Stain augmentation using stain separation improves classification performance in heterogenetic H\&E stained histopathology breast cancer dataset" medRxiv (2024): 2024-09.

\bibitem{unet}
Ronneberger, O., Fischer, P., \& Brox, T. (2015). U-net: Convolutional networks for biomedical image segmentation. In Medical image computing and computer-assisted intervention–MICCAI 2015: 18th international conference, Munich, Germany, October 5-9, 2015, proceedings, part III 18 (pp. 234-241). Springer International Publishing.

\bibitem{efficientnet}
Tan, M. (2019). Efficientnet: Rethinking model scaling for convolutional neural networks. arXiv preprint arXiv:1905.11946.

\end{thebibliography}
\end{document}